\documentclass[fleqn,twoside]{article}
\usepackage{espcrc2}

\newcommand{\be}{\begin{equation}}
\newcommand{\ee}{\end{equation}}
\newcommand{\bea}{\begin{eqnarray}}
\newcommand{\eea}{\end{eqnarray}}

\newcommand{\ms}{\medskip}

\newcommand{\n}{\noindent}
\newcommand{\bc}{\begin{center}}
\newcommand{\ec}{\end{center}}
\newcommand{\bu}{\begin{underline}}
\newcommand{\eu}{\end{underline}}

\title{Gauge-invariant quark and gluon fields in QCD:\\
dynamics, topology, and the Gribov ambiguity}

\author{Kurt Haller\address[MCSD]{Department of Physics, University of Connecticut,
        Storrs, CT 06269, USA}%
        \thanks{E-mail: khaller@uconnvm.uconn.edu}
\thanks{Research supported by the Department of Energy under Grant No.~DE-FG02-92ER40716.00.}}
      
\begin{document}

\begin{abstract}
We review the implementation, in a temporal-gauge formulation of QCD, of the non-Abelian Gauss's law and the
construction of gauge-invariant gauge and matter fields.  We then express the QCD Hamiltonian in terms of these
gauge-invariant operator-valued fields, and discuss the relation of this Hamiltonian and the gauge-invariant fields
to the corresponding quantities in a Coulomb gauge formulation of QCD. We argue that a representation of QCD 
in terms of gauge-invariant quantities could be particularly useful for understanding low-energy phenomenology.
We present the results of an investigation into the topological properties of the gauge-invariant fields, and show
that there are  Gribov copies of these gauge-invariant gauge fields, which are  constructed in the temporal gauge,
even though the conditions that give rise to Gribov copies do not obtain for the gauge-dependent
temporal-gauge fields.  
\vspace{1pc}
\end{abstract}

\maketitle

\section{INTRODUCTION}

I will review here  the construction of 
gauge invariant non-Abelian gauge and matter fields and the use
of these fields for a discussion of QCD dynamics
and of Gribov copies of gauge fields from a somewhat novel perspective. To 
illustrate one reason for our interest in
formulating QCD in terms of gauge-invariant fields, it  is helpful to first focus attention on QED. When we formulate
QED in one of a number of gauges --- for example, the Lorentz gauge or the temporal (Weyl) gauge --- and transform to
a representation in which the charged matter field and the gauge field are gauge-invariant (the former having been
obtained by use of a transformation due to Dirac,\cite{diracgauge} the
latter being just the  transverse part of the gauge field) we obtain the following 
Hamiltonian:\cite{khqedtemp,khelqed} 
\bea
{\hat H}_{QED}&&\!\!\!\!\!\!\!\!\!={\int}d{\bf r}\left[{\textstyle \frac{1}{2}}{\Pi}_i({\bf r}){\Pi}_i({\bf
r})+ {\textstyle \frac{1}{4}}F_{ij}({\bf r})F_{ij}({\bf r})\right.\nonumber \\
&&\!\!\!\!\!\!\!\!\!\left.+{\psi^\dagger}({\bf r})\left(\beta m-i\alpha_{i}\partial_{i}\right)\psi({\bf
r})-A^{(T)}_i({\bf r})j_i({\bf r})\right]\nonumber \\ &&\!\!\!\!\!\!\!\!
+{\int}d{\bf r}d{\bf r}^{\prime} \frac{j_0({\bf r})
j_0({\bf r}^{\prime})}
{8{\pi}|{\bf r}-{\bf r}^{\prime}|}+H_g\,.
\label{eq:Hqedtc}
\eea
We can recognize this transformed Hamiltonian as the sum of the Coulomb-gauge Hamiltonian 
and $H_g$, which is gauge-dependent. For the temporal gauge, $H_g$ is given by 
$$
H_g\!=\!-{\textstyle \frac{1}{2}}{\int}\!d{\bf r}\!\left(\!{\partial}_i{\Pi}_i({\bf r})\frac{1}{\partial^2}j_0({\bf
r})+ j_0({\bf r})\frac{1}{\partial^2}{\partial}_i{\Pi}_i({\bf r})\!\right)\,
\label{eq:Hg}
$$
where ${\Pi}_i({\bf r})$ is the negative electric field as well as  the momentum conjugate to $A_i({\bf r})$,
where $j_0=e\psi^{\dagger}\psi$, and $\psi$ is the gauge-invariant charged-matter field in this transformed
representation. ${\partial}_i{\Pi}_i{\approx}0$ is the form that 
Gauss's law takes in the transformed representation, with the charge density 
$j_0$ included but not appearing explicitly because a unitary transformation very much like
the one introduced in Ref.\cite{diracgauge} has folded it into ${\partial}_i{\Pi}_i$, which we therefore refer to as
the Abelian ``Gauss's law operator'' in the transformed representation. The $\approx$ indicates that the equality is 
``soft'' --- $i.\,e.$ that it is true only on a suitably defined constraint surface, or only
when applied to a set of appropriately fashioned state vectors. 
As was shown in Refs.
\cite{khqedtemp,khelqed} for a variety of gauges,
$H_g$ plays no role whatsoever in the time-evolution of state vectors, and therefore does not affect any of the
physical results obtained from the application of ${\hat H}_{QED}$. 

In the history of electrodynamics, the macroscopic long-range forces
that dominate the classical phenomenology were very familiar long before 
photon-electron scattering became an important concern. But let us imagine a fictitious scenario in which
photon-electron scattering phenomenology was our first experience with electrodynamics, and that we knew
the Lagrangian of covariant-gauge  QED and Feynman rules long before we knew about electrostatics. If, at that point,
someone had expressed that  theory in terms of gauge-invariant ``physical'' variables, and had obtained the
Hamiltonian in  Eq. (\ref{eq:Hqedtc}), it would have become
apparent that this Hamiltonian was not very appropriate for generating a renormalizable $S$-matrix.
But it would also  have become clear that, as a theory for low-energy phenomenology such as electrostatics, it was
superior to formulations that used gauge-dependent fields; that, in fact, the Coulomb interaction suffices for
understanding the energy levels and wave functions of almost all atoms, and that it is very useful in the classical
domain as well. 

One of our purposes in this work is to explore whether a similarly useful role can be assigned to QCD expressed
in terms of gauge-invariant field variables.
In order to examine this question, we have
implemented the non-Abelian Gauss's law for a temporal-gauge formulation of QCD by explicitly constructing states
$|\Psi\rangle$ that are  annihilated by the non-Abelian Gauss's law operator ${\hat {\cal G}}^a({\bf{r}})$ given by
\be
{\hat {\cal G}}^a({\bf{r}})=\overbrace{\partial_i\Pi^a_i({\bf{r}})+
\underbrace{gf^{abc}A^b_i({\bf{r}}){\Pi}^c_i({\bf{r}})}_{J^a_0({\bf{r}})}}^{D_i\Pi^a_i({\bf{r}})\,{\equiv}\,
{\cal G}^a({\bf{r}})}+j^a_0({\bf{r}}),  
\ee
where $j^a_0=g\psi^{\dagger}\frac{{\lambda}^a}{2} \psi$ is the quark
color-charge density,
$D_i\Pi^a_i$ is the ``pure glue'' form of the Gauss's law operator, and 
$J^a_0({\bf{r}})=gf^{abc}A^b_i{\Pi}^c_i$ is the  
color-charge density of the gauge field.\cite{CBH2}
\section{GAUSS'S LAW AND GAUGE INVARIANCE}
The states that are annihilated by the ``pure glue'' Gauss's law operator ${\cal G}^a({\bf{r}})$ have the 
form $|\Psi\rangle=\Psi|\Phi\rangle$, where $|\Phi\rangle$ represents a state annihilated by ${\partial}_i{\Pi}_i$,
the Abelian part of $D_i\Pi_i$, and where $\Psi$ is given by 
$\Psi={\|}\,\exp({\cal{A}})\,{\|}$.  The $||\;\;||$-ordered product orders terms so that 
all 
functionals of $A^a_i$ are 
to the left of all functionals of $\Pi^b_j\,.$  ${\cal{A}}$ is the integral operator
\be
{\cal{A}}=
i{\int}d{\bf{r}}\;\overline{{\cal{A}}_{i}^{\gamma}}({\bf{r}})\;
\Pi_i^{\gamma}({\bf{r}}),
\ee
and $\overline{{\cal{A}}_{i}^{\gamma}}({\bf{r}})$ is the {\em resolvent field}. In the course of this investigation,
it  becomes apparent that the
resolvent field is central to achieving our objective. 

In earlier work, we have obtained an integral equation for the resolvent field,\cite{CBH2} given by 
\bea
&&{\int}d{\bf r}\overline{{\cal A}_{j}^{\gamma}}({\bf r})V_{j}^{\gamma}({\bf r})=
\sum_{\eta=1}^\infty
{\textstyle\frac{ig^\eta}{\eta!}}{\int}d{\bf r}\;
\!\left\{\,\psi^{\gamma}_{(\eta)j}({\bf{r}})\,+\right.\nonumber\\
&&\left.\;\;\;\;\;\;\;\;\;\;\;\;\;\;\;f^{\vec{\alpha}\beta\gamma}_{(\eta)}\,
{\cal{M}}_{(\eta)}^{\vec{\alpha}}({\bf{r}})\,
\overline{{\cal{B}}_{(\eta) j}^{\beta}}({\bf{r}})\,\right\}\;
\!\!V_{j}^{\gamma}({\bf r})\;,
\label{eq:resfld}
\eea
where  $$\overline{{\cal Y}^{\alpha}}({\bf r})=
{\textstyle \frac{\partial_{j}}{\partial^{2}}\overline{{\cal A}_{j}^{\alpha}}({\bf r})},\;\;
{\cal{M}}_{(\eta)}^{\vec{\alpha}}({\bf{r}})
=\!\!\prod_{m=1}^\eta
\overline{{\cal Y}^{\alpha[m]}}({\bf{r}}),\;\;\mbox{and}$$ $$\overline{{\cal B}_{(\eta) i}^{\beta}}({\bf r})=
a_i^{\beta}({\bf r})+\,
\left(\,\delta_{ij}-{\textstyle\frac{\eta}{(\eta+1)}}
{\textstyle\frac{\partial_{i}\partial_{j}}{\partial^{2}}}\,\right)
\overline{{\cal A}_{i}^{\beta}}({\bf r})\,;$$
$a_i^{\beta}({\bf r})$ designates the transverse part~of the gauge field.
The fact that the resolvent field $\overline{{\cal A}_{j}^{\alpha}}({\bf r})$ 
appears in $\overline{{\cal B}_{(\eta) i}^{\beta}}({\bf r})$ and also appears in 
$\overline{{\cal Y}^{\alpha}}({\bf{r}})$, which 
is raised to all powers in ${\cal{M}}_{(\eta)}^{\vec{\alpha}}({\bf{r}})$, 
makes Eq. (\ref{eq:resfld}) a nonlinear integral equation.
$f^{\vec{\alpha}\beta\gamma}_{(\eta)}$ denotes
the chain of structure constants
\begin{eqnarray}
f^{\vec{\alpha}\beta\gamma}_{(\eta)}=&&f^{\alpha[1]\beta b[1]}\,
\,f^{b[1]\alpha[2]b[2]}\,f^{b[2]\alpha[3]b[3]}\,\times\cdots \nonumber \\
&&{\times}f^{b[\eta-2]\alpha[\eta-1]b[\eta-1]}f^{b[\eta-
1]\alpha[\eta]\gamma}
\nonumber
\end{eqnarray} 
summed over repeated indices. 
$\psi^{\gamma}_{(\eta)i}({\bf{r}})$ in Eq. (\ref{eq:resfld}) depends only on the {\em gauge-dependent} gauge field and
is understood to be an inhomogeneous source term for the nonlinear integral equation. 
Iterative expansions of the resolvent field are readily obtained and have been given.\cite{BCH1} But
our main interest will be in non-iterative representations of the resolvent field. 

The apparatus we developed for implementing Gauss's law also enables us to construct gauge-invariant 
matter (quark) and gauge (gluon) fields. The basic idea is that the complete Gauss's law operator  
${\hat {\cal G}}^a({\bf{r}})$ and the ``pure glue'' Gauss's law operator ${\cal G}^a({\bf{r}})$ are 
unitarily equivalent; and that, ${\cal{U}}_{\cal{C}}$, the unitary operator that implements the transformation
$\hat{\cal{G}}^a({\bf{r}})={\cal{U}}_{\cal{C}}\,{\cal{G}}^a({\bf{r}})\,{\cal{U}}^{-1}_{\cal{C}}$,
is given by 
$${\cal{U}}_{\cal{C}}=e^{{\cal C}_{0}}
e^{\bar {\cal C}}$$
where $${\cal C}_{0}=i\,\int d{\bf{r}}\,
{\textstyle {\cal X}^{\alpha}}({\bf r})\,j_{0}^{\alpha}({\bf r})$$
$$\mbox{and}\;\;\;\;
{\bar {\cal C}}=i\,\int d{\bf{r}}\,
\overline{{\cal Y}^{\alpha}}({\bf r})\,j_{0}^{\alpha}({\bf r}),$$
the last equation showing the role of the resolvent field in this unitary equivalence.
With this unitary equivalence, ${\cal G}^a$ can be used to represent ${\hat {\cal G}}^a$ in a 
new representation. In this {\em new} representation, the quark field $\psi$ and the current density
$g{\bar \psi}\textstyle\frac{\lambda^\alpha}{2}{\gamma^\mu}\psi$
are gauge-invariant because they commute with ${\cal G}^a$.
This unitary equivalence can then be used to construct operator-valued fields
that are  gauge invariant in the {\em original} representation:
$${\psi}_{\sf GI}({\bf{r}})={\cal{U}}_{\cal C}\,
\psi({\bf{r}})\,{\cal{U}}^{-1}_{\cal C}\,\mbox{and}\;\,
{\psi}_{\sf GI}^\dagger({\bf{r}})={\cal{U}}_{\cal C}\,
\psi^\dagger({\bf{r}})\,{\cal{U}}^{-1}_{\cal C}.$$
With the Baker-Hausdorff-Campbell theorem, we obtain
$${\psi}_{\sf GI}({\bf{r}})=V_{\cal{C}}({\bf{r}})\,\psi ({\bf{r}})\;\mbox{and}\;
{\psi}_{\sf GI}^\dagger({\bf{r}})=
\psi^\dagger({\bf{r}})\,V_{\cal{C}}^{-1}({\bf{r}}),$$ where
$$V_{\cal{C}}({\bf{r}})=
\exp\left(\,-ig{\overline{{\cal{Y}}^\alpha}}({\bf{r}})
{\textstyle\frac{\lambda^\alpha}{2}}\,\right)\,
\exp\left(-ig{\cal X}^\alpha({\bf{r}})
{\textstyle\frac{\lambda^\alpha}{2}}\right).$$
Because the commutator algebra of the $\lambda^a$ 
matrices is closed, $V_{\cal{C}}({\bf{r}})$ can be expressed as
\be
V_{\cal C}({\bf{r}})=\exp\left[-ig{\cal Z}^\alpha({\bf{r}})
{\textstyle\frac{{\lambda}^\alpha}{2}}\right],
\label{eq:zform}
\ee
 where ${\cal Z}^\alpha({\bf{r}})$ is 
a functional of ${\cal X}^\alpha({\bf{r}})$ and ${\overline{{\cal{Y}}^\alpha}}({\bf{r}})$; 
in the SU(2) case, the relation among these quantities is that of angles in rigid-body rotations. 
In the form given by Eq. (\ref{eq:zform}), $V_{\cal C}({\bf{r}})$
has the formal structure of an operator that gauge-transforms a  
charged field. However, ${\cal Z}^\alpha$, which would have to be a $c$-number valued field
for $V_{\cal C}$ to be such a gauge-transformation, in fact is operator-valued;
and under a gauge transformation which transforms the matter field by the $c$-number function
${\omega}^{\gamma}({\bf{r}})$, the matter field and $V_{\cal C}$ transform as
$${\psi}{\rightarrow}\exp(-i{\omega}^{\gamma}\,
\textstyle{\frac{{\lambda}^{\gamma}}{2}})\psi\;\;\;
\mbox{and}\;\;\;V_{\cal C}{\rightarrow}V_{\cal C}\exp(i{\omega}^{\gamma}\,
\textstyle{\frac{{\lambda}^{\gamma}}{2}})$$
so that $V_{\cal{C}}({\bf{r}}) \psi$ remains gauge-invariant.
Exploiting the formal similarity of the structure of the gauge-invariant matter field to a 
gauge transformation of that field enables us to also construct gauge-invariant 
gauge fields in the form~\cite{LMNP} 
$$
{\sf A}_{{\sf GI}\,i}({\bf{r}})
=V_{\cal{C}}({\bf{r}})\,{\sf A}_i({\bf{r}})\,
V_{\cal{C}}^{-1}({\bf{r}})\,
+{\textstyle\frac{i}{g}}\,V_{\cal{C}}({\bf{r}})\,
\partial_{i}V_{\cal{C}}^{-1}({\bf  r})\;,
$$
where ${{\sf A}\,i}({\bf{r}})=A_i^{b}({\bf{r}})\,{\textstyle\frac{\lambda^b}{2}}$ or, equivalently,
\be
A_{{\sf GI}\,i}^{b}({\bf{r}})\,=
A\,_{T\,i}^b ({\bf{r}}) +
[\delta_{ij}-{\textstyle\frac{\partial_{i}\partial_j}
{\partial^2}}]\,\overline{{\cal A}_{i}^b} ({\bf{r}})\;.
\ee
We can take this formal similarity further, by noting that for ${\sf A}_{\,0}({\bf{r}})=0$, 
\be
{\sf A}_{{\sf GI}\,0}({\bf{r}})={\textstyle\frac{i}{g}}\,V_{\cal{C}}({\bf{r}})\,
\partial_{0}V_{\cal{C}}^{-1}({\bf  r})\,.
\ee
With these results, we can identify the gauge-invariant negative chromoelectric field as~\cite{CHQC}
\be
{\Pi}_{{\sf GI}\,i}^d={\textstyle\frac{1}{2}}{\sf Tr}[V_{\cal{C}}^{-1}
{\lambda^d}V_{\cal{C}}{\lambda^b}]
\Pi^{b}_{i}.
\ee
Finally, the gauge-invariant chromomagnetic field is 
$$F_{{\sf GI}\,ij}^{a}=\partial_jA_{{\sf GI}\,i}^{a}-\partial_iA_{{\sf GI}\,j}^{a}-
g\epsilon^{abc}A_{{\sf GI}\,i}^{b}A_{{\sf GI}\,j}^{c}.$$
\section{GAUGE-INVARIANT QCD\\ DYNAMICS}
In this section, we will make use of earlier work, in which the 
temporal-gauge Hamiltonian was expressed {\em entirely} in terms of
the gauge-invariant quantities that we introduced in earlier sections.\cite{CBH2,BCH3,HGrib}
In this form, the Hamiltonian is
\bea
\,{\hat H}_{\sf GI}\!\!\!\!&=&\!\!\!\!\!\!\int\! \!d{\bf r}\left[\!\! \ {\textstyle \frac{1}{2}}
\Pi^{a\,{\dagger}}_{{\sf GI}\,i}({\bf r})\Pi^{a}_{{\sf GI}\,i}({\bf r})
+  {\textstyle \frac{1}{4}} F_{{\sf GI}\,ij}^{a}({\bf r}) F_{{\sf GI}\,ij}^{a}({\bf r})\right.\nonumber\\
&+&\left.{\psi^\dagger}({\bf r})\left(\beta m-i\alpha_{i}
\partial_{i}\right)\psi({\bf r})\right] + \tilde{H}^{\prime}+H_{\cal G}
\label{eq:HQCDN}
\eea
with
\bea
\,\tilde{H}^{\prime}\!\!\!\!\!&=&\!\!\!\!\!\!\int\!\! d{\bf r}\left({\textstyle\frac{1}{2}}J_{0\,({\sf
GI})}^{a\,\dagger}({\bf r}) 
\frac{1}{\partial^2}{\cal K}_0^a({\bf r})+
{\textstyle\frac{1}{2}}\,{\cal K}_0^a({\bf r})\frac{1}{\partial^2}J_{0\,({\sf GI})}^{a}\right.\nonumber\\
&-&\left.{\textstyle\frac{1}{2}}{\cal K}_0^a({\bf r})\frac{1}{\partial^2}{\cal K}_0^a({\bf r})-
j^a_i({\bf r})A^a_{{\sf GI}\,i}({\bf{r}})\right)
\label{eq:Hprime}
\eea
and
\be
\,H_{\cal G}=-{\textstyle\frac{1}{2}}\int\!\!d{\bf r}\left[{\cal G}_{\sf GI}^{a}
\frac{1}{\partial^2}{\cal K}_0^a({\bf r})+
{\cal K}_0^a({\bf r})\frac{1}{\partial^2}{\cal G}_{\sf GI}^{a}\right].\;\;\;
\label{eq:HamGauss}
\ee
${\cal K}_0^a$ describes a gauge-invariant nonlocal ``effective'' quark color-charge density, which is related to the
local (but also gauge-invariant) quark color-charge density by a Faddeev-Popov equation, as shown by
\be
{\cal K}_0^a+g{\epsilon}^{avb}A_{{\sf GI}\,i}{\textstyle}{\frac{\partial_i}{\partial^2}}{\cal K}_0^b=-j^a_0.
\label{eq:Kscr}
\ee
and $J_{0\,({\sf GI})}^{a}$ is the gauge-invariant ``glue'' color-charge density
$J_{0\,({\sf GI})}^{a}=gf^{abc}A_{{\sf GI}\,i}^{b}{\Pi}_{{\sf GI}\,i}^c$. 
We observe that $\tilde{H}^{\prime}$ manifests interesting similarities to the QED Hamiltonian shown in 
Eq. (\ref{eq:Hqedtc}). One of its terms describes the
interaction of the gauge-invariant (transverse) gauge field with the transverse color-current density, which is
also gauge-invariant, and, as is true for a cognate term in QED,  not likely to make important contributions at
low energies. $\tilde{H}^{\prime}$ also contains terms  describing Coulomb interactions between gauge-invariant
quark-quark and quark-gluon color-charge densities. 

${\hat H}_{\sf GI}$  has some features in common with expressions obtained by other investigators who have formulated
QCD in the Coulomb gauge.\cite{Coul} But ${\hat H}_{\sf GI}$ 
also differs from Hamiltonians in Coulomb-gauge formulations of QCD in a
number of ways, for example in the presence of
$H_{\cal G}$, which is the term that ``remembers'' that this formulation is specific to the temporal gauge, but
which, as was  shown in Ref.\cite{HGrib}, cannot affect any of the physical consequences obtained with ${\hat H}_{\sf
GI}$. The gauge-invariant fields resemble those of the Coulomb gauge, and have equal-time commutation rules very much
like those obtained by Schwinger for that gauge,\cite{schwingerb} but differ from them in operator order.
The situation in QCD is therefore very  similar to the one we described for QED in connection with Eq.
(\ref{eq:Hqedtc}). The nonlocality of ${\cal K}_0^a$, and its interactions with itself and with the gauge-invariant 
gluon color-charge density, provide an incentive to examine the long-range properties of the interaction described
by $\tilde{H}^{\prime}$ --- in particular, whether it might describe a confining force acting on color-bearing
objects. In Ref.\cite{CHQC}, we also argued that in the regime in which QCD variables describe hadronic interactions,
the  form of ${\cal K}_0^a$ is suggestive of color transparency for combinations of quarks in a color-singlet 
configuration. 

Finally, it is also worth noting that, in the transformation to a representation in terms of gauge-invariant fields,
Faddeev-Popov ghosts have not been introduced into the QCD Hamiltonian. Our procedure for arriving at a 
representation of QCD in terms of ``physical'' fields does not require the introduction of 
Faddeev-Popov ghost fields. 
\section{TOPOLOGY AND GRIBOV COPIES}
In this section, I will review investigations into the topological properties of the resolvent field
for two-color QCD, in which the Pauli spin matrices $\tau^a$ replace the Gell-Mann matrices $\lambda^a$.
By representing the resolvent field $\overline{{\cal{A}}_{i}^{\gamma}}({\bf{r}})$ as a function of spatial 
variables that are second-rank tensors in the combined spatial and SU(2) indices, we have obtained and 
solved a nonlinear differential equation for ${\overline{\cal{N}}}$,\cite{HCC} which is related to the resolvent
field by 
\be
{\overline{\cal{N}}}=\left(\textstyle{\frac{\partial_i}{\partial^2}}\overline{{\cal{A}}_{i}^{\gamma}}
\textstyle{\frac{\partial_i}{\partial^2}}\overline{{\cal{A}}_{i}^{\gamma}}\right)^{1/2}.
\label{eq:nbar}
\ee
This equation, 
\begin{eqnarray}
\frac{d^2\,\overline{\cal{N}}}{du^2}\!\!\!\!&+&\!\!\!\!\frac{d\,\overline{\cal{N}}}{du}+
2\left[{\cal{N}}{\cos}(\overline{\cal{N}}+{\cal{N}})-
{\sin}(\overline{\cal{N}}+{\cal{N}})\right]\nonumber \\
&+&2gr_0\exp(u)\left\{{\cal T}_A\left[{\cos}(\overline{\cal{N}}+{\cal{N}})-1\right]\right.\nonumber\\
&&-\left.{\cal T}_C\,
{\sin}(\overline{\cal{N}}+{\cal{N}})\right\}=0
\label{eq:nlde}
\end{eqnarray}
where $u=\ln(r/r_0)$, can also describe a driven, damped, pendulum with the important proviso that 
${\overline{\cal{N}}}$, given in Eq. (\ref{eq:nbar}), must be bounded in $u$ in the entire interval
$(-\infty,\infty)$, whereas the pendulum equation only applies to the interval $(0,\infty)$. In Ref.\cite{HCC}, we
graphically  display numerical solutions of Eq. (\ref{eq:nlde}) and show that, for the same choice of ``source''
terms, there are a number of bounded solutions in the interval $(-\infty,\infty)$ 
and that these not only differ from each other, but that they also can have different asymptotic values as
$u\rightarrow\infty$; and that different asymptotic values of ${\overline{\cal{N}}}$, in that limit, correspond to a
variety of winding numbers, many half-integer valued, or fractional valued. 
We also pointed out in this work that the solutions of Eq. (\ref{eq:nlde}), and the asymptotic limits of these 
solutions as $u\rightarrow\infty$, are not strongly dependent on the functional forms of the source terms
${\cal{N}}$, ${\cal T}_A$, and ${\cal T}_C$.
Eq. (\ref{eq:nlde}) is of the general
form of an equation given by Gribov to document the existence of multiple copies of 
Coulomb-gauge fields,\cite{gribovb}
$$\frac{d^2\,\phi}{du^2}+\frac{d\,\phi}{du}-2{\sin}(\phi)\left(1-f(u)\right)=0,$$
and the multiple solutions for $\overline{{\cal{A}}_{i}^{\gamma}}({\bf{r}})$ correspond to Gribov copies of 
the gauge-invariant gauge field. This fact has led us to the following observations about Gribov copies
in the temporal and Coulomb gauges.\ms

\n
$\bullet$ When QCD is quantized in the Coulomb gauge, 
the quantization procedure is impeded by the existence of Gribov copies, 
because the non-uniqueness of the inverse of the Faddeev-Popov 
operator prevents the inversion of the Dirac constraint
commutator matrix.\cite{ramond}\ms

\n
$\bullet$ When the quantization is carried out in the temporal 
gauge (or another algebraic gauge) no impediments to
the inversion of the commutator matrix arise, and the 
procedure can be carried out consistently, without any concern about nonunique inverses of that matrix.
It is in this sense that the statement that there are no Gribov 
copies of the {\em gauge-dependent} temporal-gauge field can be understood.\cite{wein} 
But, in contrast to the Coulomb gauge (or in other gauges in which Gauss's law is a secondary constraint) 
Gauss's law remains to be implemented after quantum rules have been imposed on the operator-valued 
temporal-gauge fields.\cite{jackiw}\ms

\n
$\bullet$ Gribov copies arise in the temporal gauge when Gauss's law is implemented,
and they are a feature of the {\em gauge-invariant,} but not the {\em gauge-dependent} fields. 
It is the imposition
of gauge invariance that produces gauge fields that have Gribov copies in QCD. This is consistent with a proof 
given by Singer,  and with remarks in his paper about the absence of Gribov copies in axial gauge
formulations in which $n^{\nu}A^a_\nu=0$ defines the gauge.\cite{Singer}


\newpage



\begin{thebibliography}{99}
\bibitem{diracgauge} P. A. M. Dirac, Canad. J. Phys. {\bf 33} (1955) 650.
\bibitem{khqedtemp}K. Haller, Phys.\ Rev. {\bf D 36} (1987) 1830.
\bibitem{khelqed}K. Haller and E. Lim-Lombridas, Foundations of Physics {\bf 24} (1994) 217.
\bibitem{CBH2}L. Chen, M. Belloni and K. Haller, Phys.\ Rev.
{\bf D 55} (1997) 2347.
\bibitem{BCH1} M. Belloni, L. Chen and K. Haller, Phys.\ Lett. {\bf B373}, (1996) 185.
\bibitem{LMNP}Iterative expansions of these expressions agree with the perturbative results given in
M. Lavelle and D. McMullan, Phys.\ Lett.\ {\bf B 329} (1994) 68;~Phys. Rept. {\bf 279} (1997) 1.
\bibitem{CHQC}L. Chen and K. Haller, Int. J. Mod. Phys. {\bf A 14} (1999) 2745.
\bibitem{BCH3} M. Belloni, L. Chen and K. Haller, Phys.\ Lett.
{\bf B 403} (1997) 316.
\bibitem{HGrib}K. Haller, Int. J. Mod. Phys. {\bf A16}2 (2001) 2789.
\bibitem{Coul}J. Schwinger, Phys. Rev. {\bf 125} (1962) 1043; 
N. H. Christ and T. D. Lee, Phys. Rev. {\bf D 22}
(1980) 939; M. Creutz, I. J. Muzinich, and T. N. Tudron, Phys. Rev. {\bf D 19} (1979) 531;
J. L. Gervais and B. Sakita, Phys. Rev. {\bf D 18} (1978) 453.
\bibitem{schwingerb}J. Schwinger, Phys. Rev. {\bf 127} (1962) 324.
\bibitem{HCC}K. Haller, L. Chen and Y. S. Choi, Phys. Rev. {\bf D 60} (1999) 125010.
\bibitem{gribovb}V. N. Gribov, {\em Instability of non-Abelian gauge theories and 
impossibility of choice of Coulomb gauge,} 
Lecture at the 12th Winter School of the Leningrad Nuclear Physics Institute (unpublished).
\bibitem{ramond}For example, P. Ramond, {\em Field Theory, a Modern Primer} Second Edition (Addison Wesley,
New York, 1990).
\bibitem{wein}For example, S. Weinberg, {\em The Quantum Theory of Fields} Vol. II (Cambridge Univ. Press, 
Cambridge, UK, 1995), Section 15.4. 
\bibitem{jackiw}R. Jackiw, Rev.\ Mod.\ Phys. {\bf 52} (1980) 661.
\bibitem{Singer}I.~M.~Singer, Commun.~Math.~Phys. {\bf 60} (1978) 7.



\end{thebibliography}
\end{document}